\documentclass{aa} 

\usepackage{epsfig}
\usepackage{graphicx}
\usepackage{natbib}
\usepackage{txfonts}
\begin{document}
   \title{Estimation of Galactic model parameters in high latitudes with 2MASS}

   \author{A. Cabrera-Lavers\inst{1,2}, S. Bilir\inst{3},
           S. Ak\inst{3}, E. Yaz\inst{3},
	  \and
          M. L\'opez-Corredoira\inst{1}
          }

   \offprints{antonio.cabrera@gtc.iac.es}

   \institute{Instituto de Astrof\'{\i}sica de Canarias, E-38205 La Laguna, Tenerife, Spain\\
            \and
	    GTC Project Office, E-38205 La Laguna, Tenerife, Spain\\
                       \and
	 Istanbul University Science Faculty, Department of Astronomy and Space
	 Sciences, 34119, University-Istanbul, Turkey\\
                  }

   \date{Received XX; accepted XX}

% \abstract{}{}{}{}{}
% 5 {} token are mandatory

 \abstract
   {In general, studies focused on the Milky Way's structure show a range of values 
   when deriving different Galactic parameters such as radial scalelengths, vertical scaleheights, or local space densities.
   Those values are also dependent on the Galactic coordinates under consideration for the corresponding analysis, as a direct
   consequence of be observing a structure (Our Galaxy), that is far from being as smooth and well-behaved as models usually consider.}
   {In this paper, we try to find any dependence of the Galactic structural parameters with the Galactic longitude 
   for either the thin disc or the thick disc of the Milky Way, that would indicate to 
   possible inhomogeneities or asymmetries in those Galactic components.}
   {Galactic model parameters have been estimated for a set of 36 high-latitude fields 
    with Two Micron All Sky Survey (\emph{2MASS}) photometry. Possible variations with the Galactic longitude
    of either the scaleheight and the local space density of these components are explored.}
   {Galactic model parameters for the different fields show that, effectively,
they are Galactic longitude-dependent. The thick disc scaleheight
changes from $\sim$800 pc at $150^{\circ}<l<230^{\circ}$ to $\sim$1050 pc at
$|l|<30^{\circ}$. A plausible explanation for this finding might be the effect
of the flare in this Galactic component that changes the scaleheight ($h_z$)
with Galactocentric distance ($R$) following the approximate law:
$h_z (R)= (940\pm20)\times [1 - (0.12\pm0.02) (R-R_\odot)]$. The effect
of the flare is more stressed in some lines of sight than in others, producing
the observed changes in the parameters with the Galactic coordinates
used to derive them.}
   {}

\authorrunning{Cabrera-Lavers et al.}
\titlerunning{Galactic model parameters}

   \maketitle
%
%________________________________________________________________

\section{Introduction}
Detailed study of star counts in the Milky Way permit us to recover basic
structural parameters, such as its disc scalelength and disc scaleheight. Our knowledge of 
the structure of the Milky Way, as inferred from star count data, is about to enter the next level of 
precision with the advent of new surveys such as {\em 2MASS\/}. This will require the refinement of models of Galactic 
structure to fit the parameters of the basic components of Milky Way.

Researchers have used different methods to determine the Galactic model parameters. In Table 1 of \cite{KBH04} we can
found an exhaustive summary of the different values obtained for the sctructural parameters of the discs and halo of 
the Milky Way. One can see that there is an evolution for the numerical values of model parameters. The local 
space density and the scaleheight of the thick disc can be given as an example. The evaluations of the thick disc have steadily
 moved towards shorter scaleheights, from 1.45 to 0.65 kpc \citep{GR83, Chen01} and higher local densities (2–-10 per cent). 
 In many studies the range of values for the parameters is large. For example, \cite{Chen01} and \cite{Siegel02} give 6.5–-13 and 
 6-–10 per cent, respectively, for the local space density for the thick disc.

Different model parameters revealed for the Galactic disc may due to the warp and flare. The disc of the Milky Way is far 
from being radially smooth and uniform. On the contrary, it presents strong asymmetries in its overall shape. While the warp 
bends the Galactic plane upwards in the first and second Galactic longitude quadrants (0$^\circ\le l \le180^\circ$) and downwards 
in the third and fourth quadrants (180$^\circ\le l \le360^\circ$), the flare changes the scaleheight as a function of radial distance.

The warp is present in all Galactic components: dust \citep{DS01, Mar06}, gas \citep{Burton88, DS01, NS03, Levine06, VB06}, and 
stars \citep{LBB02, Momany06}. All these components have the same node position and their distributions are asymmetric. However, 
the amplitude of the warp seems to depend slightly on the component one looks at: the dust warp seems to be less pronounced than 
the stellar and gaseous warps, that share approximately the same amplitude \citep{LBB02, Momany06}. 

The stellar and gaseous flaring for the Milky Way are also compatible \citep{Momany06}, showing that the scaleheight 
increases with the Galactocentric radius for $R>5$ kpc \citep{Kent91, DS01, NJ02, L02, Momany06}. The behaviour of
 this flare in the central discs of spiral galaxies is not so well studied due to inherent difficulties in separating 
 the several contributions to the observed counts or flux. \cite{L04}, for example, found that there is a deficit of stars 
 respect to the predictions of a pure exponential law in the inner 4 kpc of the Milky Way, that could be explained as being a 
 flare which displaces the stars to higher heights above the plane as we move to the Galactic center.

In this scenario, where on the one hand the mean disc ($z=0$) can be displaced as much as 2 kpc between the location of the maximum and the
minimum amplitude of the warp \citep{DS01, L02, Momany06}, and on the other hand the scaleheight of the stars can show differences up to 50
per cent of the value for $h_z(R_{\odot})$ in the range $5 < R < 10$ kpc \citep{Al00, L02, Momany06} to fit a global Galactic disc model which
 accounts for all these inhomogeneities is, at least, tricky. It is because of this than the results in the Galactic model parameters might depend on the sample of Galactic coordinates used, 
 as the combined effect of the warp and flare will be different at different directions in the Galaxy, hence at different lines of sight.

In this paper, we derived  the structural parameters of the thin and thick discs of the Milky Way by using data from
 {\em 2MASS\/} survey, to observe changes in the parameters with the Galactic longitude. We briefly describe the {\em 2MASS\/} data, 
  density functions and the estimation of the Galactic model parameters in Section 2. Dependence of the Galactic model parameters 
  with the Galactic longitude is given in Section 3. Finally, our main results are discussed and summarized in Sections 4 and 5, respectively.

\section{2MASS data}
The \emph{Two Micron All Sky Survey} \citep[{\em 2MASS},][]{2mass06} provides the most complete database of near infrared (NIR) 
Galactic point sources available to date. During the development of this survey two highly-automated 1.3-m telescopes were used,
 one at Mt. Hopkins (AZ, USA) to observe the Northern Sky and one at Cerro Tololo Observatory (CTIO, Chile) to complete the Southern 
 counterpart of the survey. Observations cover approximately 97 per cent of the Sky with simultaneous detections in $J$ (1.25$\mu$m), 
 $H$ (1.65 $\mu$m), and $K_s$ (2.17 $\mu$m) bands up to limiting magnitudes of 15.8, 15.1, and 14.3, respectively.

In this paper we have used data from the \emph{All Sky Release} of {\em 2MASS}, made available to the public on March 2003, 
that includes a point source catalogue of $\sim$470 million stars \citep{Cu03}. We extracted those areas at Galactic latitudes 
of $45^\circ \le b \le55^\circ$ and $60^\circ \le b \le 70 ^\circ$, where both the dif\-fe\-ren\-tial reliability and completeness of 
the {\em 2MASS} catalogue are 0.99, thus the nominal limiting magnitudes of the survey can be easily achieved.

\subsection{Stellar density from the red clump population}
\label{2mass}
In \cite{L02} was presented a method of deriving stellar densities and the interstellar 
extinction along a given line of sight, which was developed in a subsequent series of papers \citep{Drimmel03, Picaud03, L04}. 
The method relies in ex\-trac\-ting the red clump (RC) population from the colour-magnitude diagrams (CMDs) as they are the dominant 
giant population \citep{Co00, H00}. These stars can be easily isolated as they form a conspicuous feature in the CMDs. The absolute 
magnitude and intrinsic colour of this population are well defined: $M(K)=-1.62\pm0.03$, $(J-K)_0=0.61\pm0.01$ with a small dependence 
with the metallicity and age \citep{Al00, GS02, Salaris02, Piet03}. Of course, there is some dispersion with respect to those values 
($\sigma [(J-K)_0]\sim 0.1$), but we know that this population is dominant and the dispersion of absolute magnitudes and colours is not very large so the use of an 
average value of $M(K)$ and $(J-K)_0$ is a good approximation. Therefore, we can extract spatial information directly from the apparent 
magnitudes and $(J-K)$ colours of the RC stars in the CMDs.

\begin{figure}[!h]
\begin{center}
\includegraphics[angle=0, width=80.2mm, height=75.0mm]{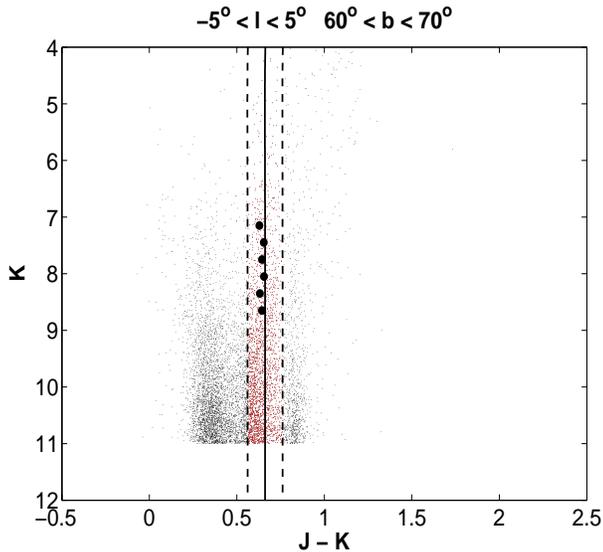}
\caption[] {$K$ vs. $J-K$ {\em 2MASS} colour-magnitude diagram for the field centred at $l=0^\circ$. 
The fitted trace that we assign to RC giants population (solid line) and the limits for extraction of the stars (dashed lines) 
are also shown (see Sect. 2 in \cite{Cabrera05} for details).} 
\label{traza}
\end{center}
\end {figure}

The method has been used before in \citet*{Cabrera05} to analyze the thick disc of the Milky Way, ex\-trac\-ting some results about the structural
parameters of this component that were in good agreement with previous estimates for the Galactic thick disc. The full details of the
application of the method can be found in that paper and also in \cite{L02} so they will not be repeated here. 
As a brief summary, the RC stars are firstly isolated in the CMDs by means of theoretical traces predicted using the "SKY" 
model \citep{SKY92}, that define the approximate area in the CMD where the RC population lies. Once these stars are identified, 
RC stars are extracted around a trace fitted to the maxima of a series of consecutive Gaussian fits to different colour histograms 
at fixed apparent magnitudes, $m_K$ (see Fig. \ref{traza}). The distance along the line of sight to those stars can be derived ea\-si\-ly 
from the apparent colour and magnitude. By obtaining the number of RC stars in each interval of apparent magnitude, the 
stellar density can be derived once the absolute magnitude and intrinsic colour of the RC are known (that is the main assumption of the 
method). As we obtain densities along the line of sight, they can be transformed into densities in cylindrical coordinates $(R,z)$.

The possible contamination due to dwarf stars in the extracted counts is the more restrictive issue 
when applying the method. To minimize this effect, we extracted only stars up to $m_K<$10 from the CMDs. In this range, 
contamination is less than 10 per cent of the total number stars, and it is even lower for brighter apparent magnitudes 
\citep{Cabrera05}. A magnitude $m_K=10$ corresponds to a distance from the Sun of approximately 2.5 kpc for a RC star, suitable for
a\-na\-ly\-zing the disc more than 1 kpc above the Galactic plane but un\-able to reach the range of distances where the halo dominates the counts. Thus, the following analysis is made ignoring this Galactic component. 

Other aspects that affect the method, as the metallicity dependence of the colour of the RC population, any possible contamination of 
lower luminosity giants in the counts, or the Malmquist bias \citep{M20} in the absolute magnitude of the RC stars have been taken into account, and their possible effects in the results were deeply discussed in Sect. 3 of \cite{Cabrera05}. In any case, the uncertainties are always lower than the one coming from the contamination of dwarf stars.

We have first collected 10 $\times$ 10 degree fields in the {\em 2MASS} catalogue centered at fixed Galactic longitudes
 (in $\Delta l=10^{\circ}$ bins), with 60$^{\circ}\leq b \leq70^{\circ}$. We have to average several fields
  to obtain a sufficient number of RC stars to build up a conspicuous feature in the CMD which could be 
   easily identifiable. The densities extracted are fitted by a double exponential following 
   eq. (\ref{ec1}), for both the thin and thick discs, with the additional constraint of producing the same local disc space
   density for the thin disc as it was obtained previously in \citet*{Cabrera05}. 

Therefore, we assume for the discs a distribution as follows:   
\begin{equation}
\rho_{i}(R,z)=n_{i}~exp(-|z|/h_{z,i})~exp(-(R-R_{0})/H_{i}),\\
\label{ec1}
\end{equation}
 
where $z=z_{\odot}+r\sin(b)$, $r$ is the distance to the object from the Sun, $b$ is the Galactic latitude, $z_{\odot}$ is 
the vertical distance of the Sun from the Galactic plane we assume to be 15 pc \citep{H94}, $R$ is the projected of the Galactocentric 
distance on the Galactic plane, $R_{0}$ is the solar distance from the Galactic center \citep[8 kpc]{R93}, 
$h_{z,i}$ and $H_{i}$ are the scaleheight and scalelength, respectively, and $n_{i}$ is the nor\-ma\-li\-zed density at the solar radius. 
The suffix $i$ takes the values 1 and 2 as long as the thin and thick discs are considered. 
   
As this study focuses on the dependence of the scaleheight and solar normalization on the Galactic longitude, for the 
    scalelengths of the thin and thick discs we used the values of 2.1 kpc \citep{L02} and 3 kpc \citep{Cabrera05} respectively, 
    which were also obtained with {\em 2MASS} data. In any case, it has to be noted that 
    the range of Galactocentric distances covered along each line of sight is so small that the 
    effect of changing the value of the radial scalelengths is negligible \citep{Cabrera05}. The possible implications of a different
    scalelength of the thick disc will be discussed in Section \ref{seccH}.

\begin{figure}
\begin{center}
\includegraphics[angle=0, width=70mm, height=55.2mm]{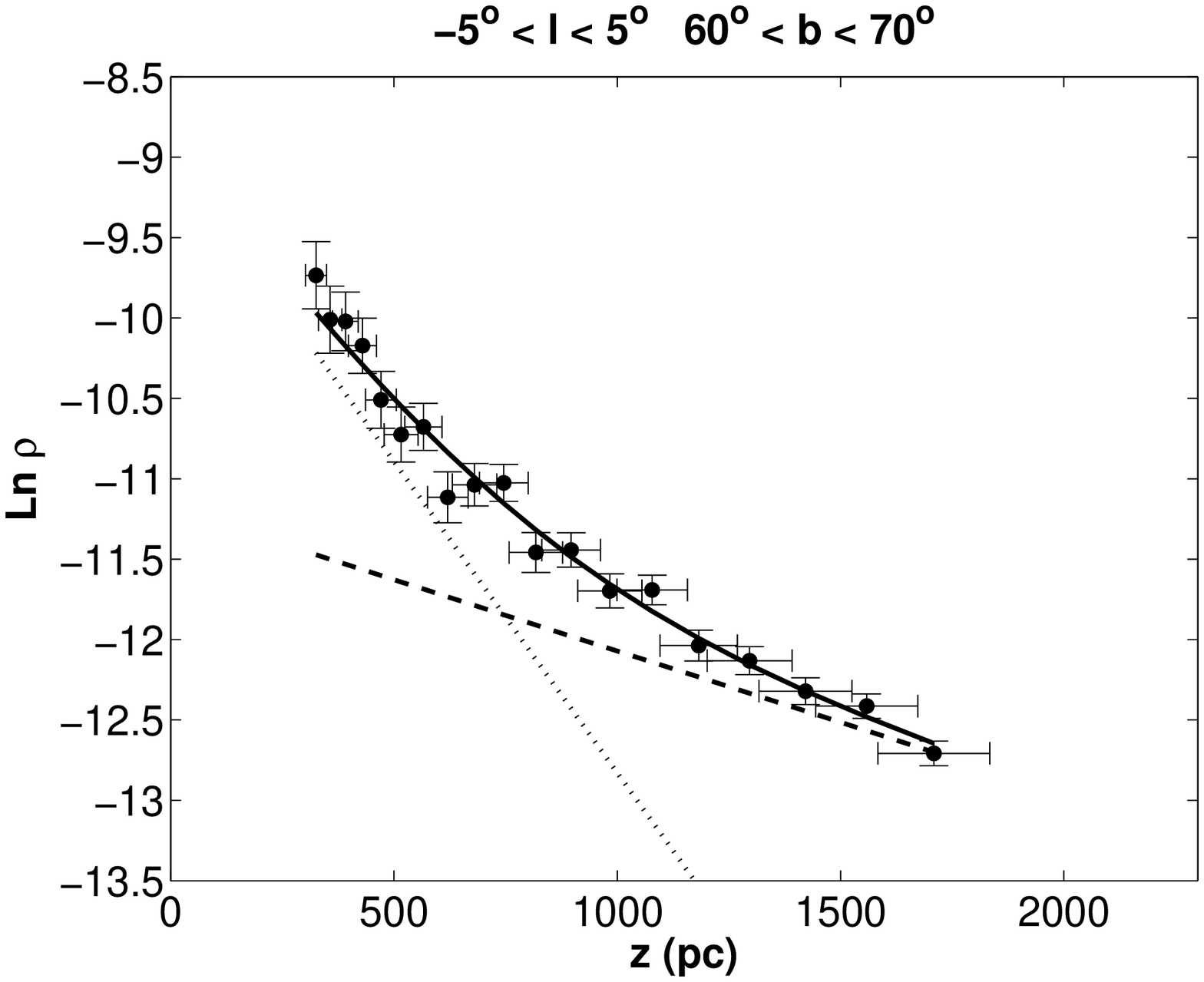}
\includegraphics[angle=0, width=70mm, height=55.2mm]{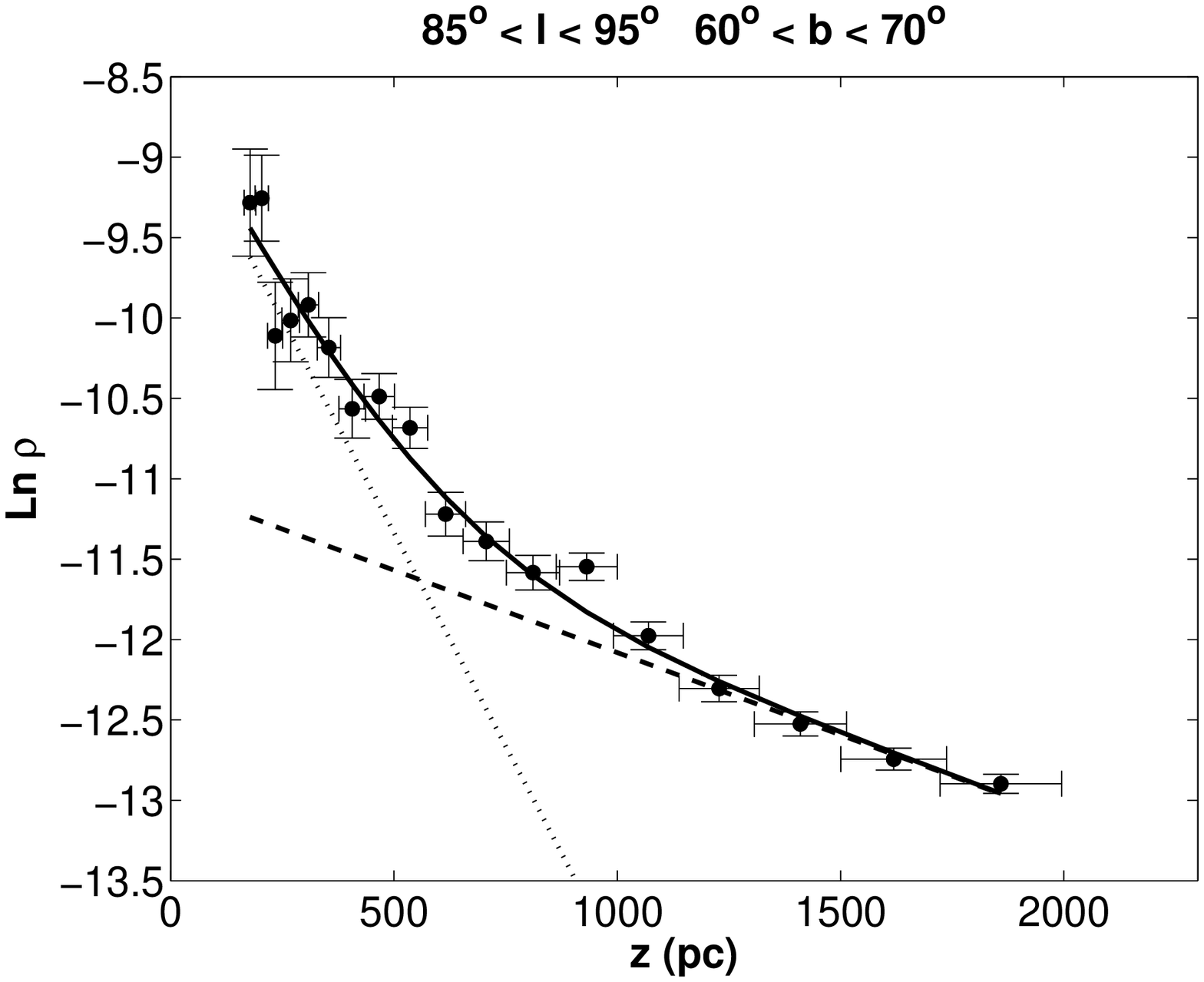}
\includegraphics[angle=0, width=70mm, height=55.2mm]{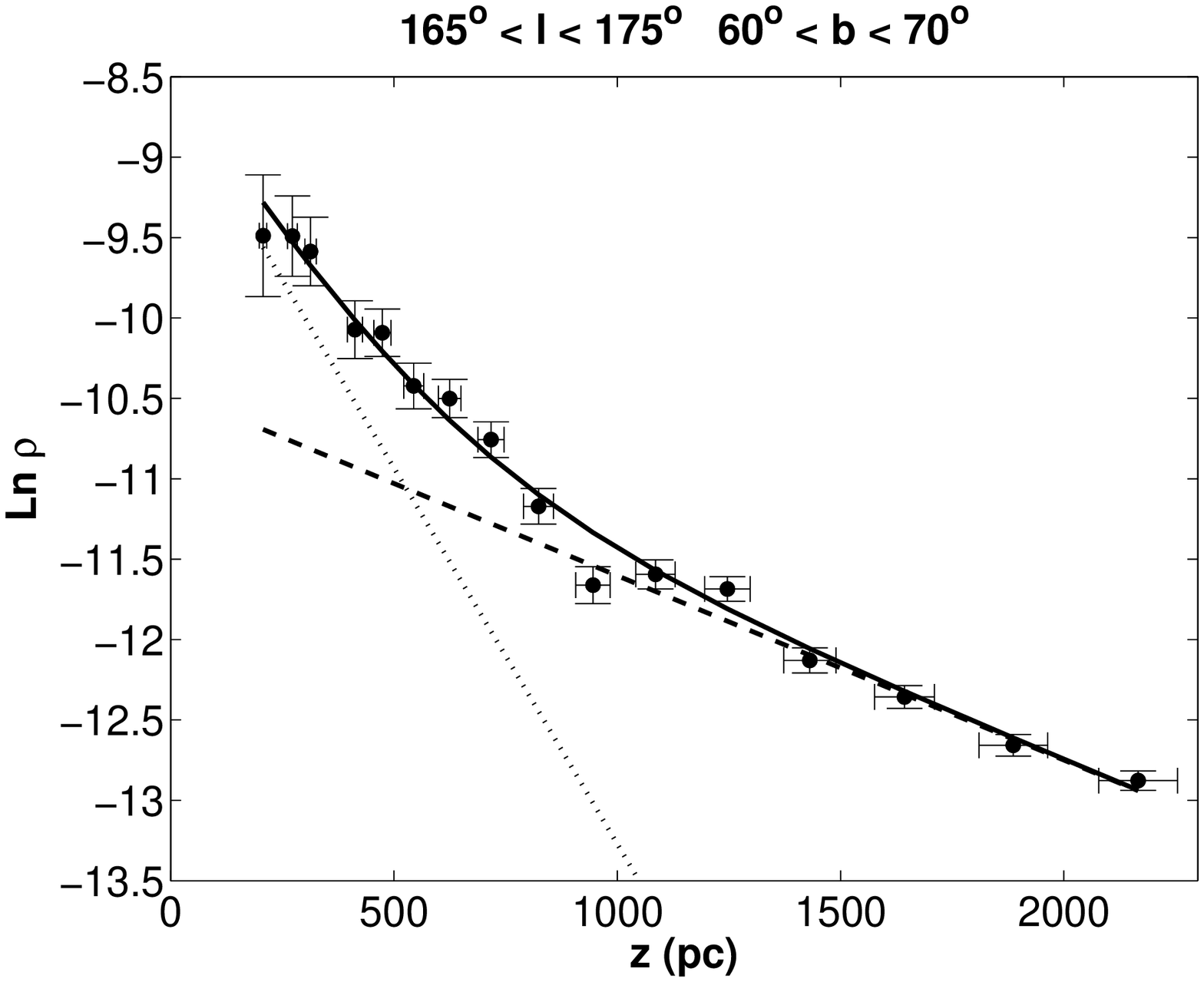}
\includegraphics[angle=0, width=70mm, height=55.2mm]{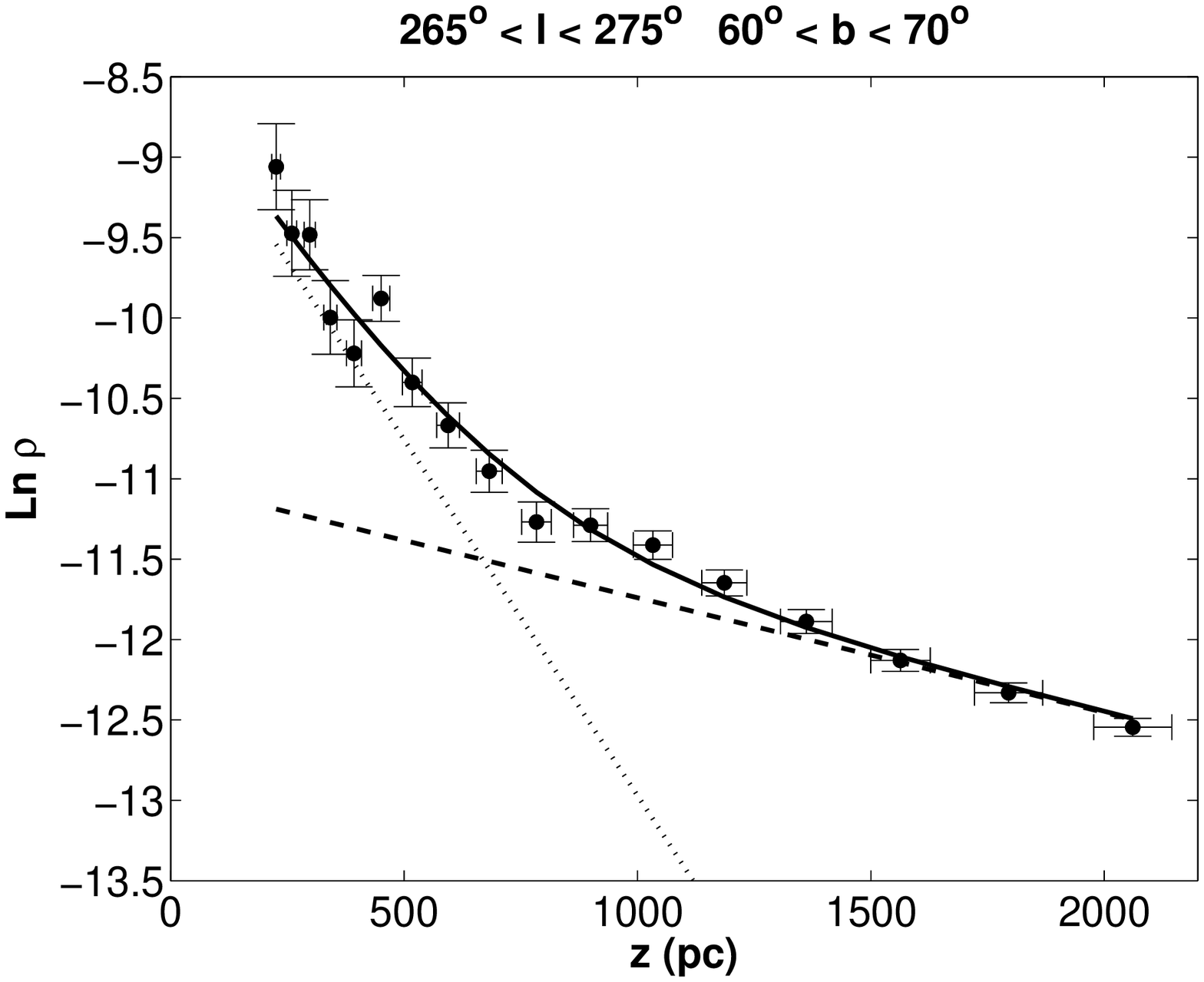}
\caption[] {Density of red clump stars (in stars pc$^{-3}$) obtained from the CMDs for the Galactic longitudes of 
 0$^\circ$, 90$^\circ$, 170$^\circ$, 270$^\circ$. The best fits for a sum of thick disc 
(dashed line) and thin disc (dotted line) components are also shown as a solid line. Error bars come from the 
Poisson noise in the extracted counts and from the uncertainty in the distance above the plane.} 
\label{dens}
\end{center}
\end {figure}

Fig. \ref{dens} shows stellar densities profiles extracted at four different Galactic longitudes, as an example of the application 
of the method, while the Galactic model parameters are given in Table \ref{table2mass}. Columns (2) and (3) give 
the scaleheight of the thin and thick discs, respectively, while the local space density of the thick disc $(n_{2}/n_{1})$ is 
given in Col. (4). Error bars in the scaleheights come from the uncertainty in the double exponential fit, by assuming Poisson 
noise in the extracted counts and  taking into account the uncertainty in the distance above the plane as we have grouped 
10 $\times$ 10 degree fields. 

\begin{table}
\center
\caption{Galactic model parameters for the 36 star fields analyzed with \emph{2MASS} data.}
\label{table2mass}
\begin{tabular}{cccc}
\hline
    & Thin disc & \multicolumn{2}{c}{Thick disc}\\
\hline
$<l>$ ($^\circ$) &  $h_{z}$ (pc)    & $h_{z}$ (pc) & $n_{2}/n_{1}~(\%)$  \\
\hline
 0    &    258$\pm$25 & 1127$\pm$142 &10.79$\pm$2.71\\
10    &    209$\pm$18 & 1043$\pm$77  &13.75$\pm$4.17\\
20    &    187$\pm$13 &  974$\pm$48  &14.48$\pm$3.62\\
30    &    161$\pm$26 &  963$\pm$101 &11.85$\pm$2.95\\
40    &    169$\pm$11 & 1074$\pm$70  & 9.63$\pm$2.41\\
50    &    197$\pm$ 7 &  909$\pm$38  & 8.08$\pm$2.02\\
60    &    202$\pm$22 &  943$\pm$90  & 9.21$\pm$2.31\\
70    &    185$\pm$13 &  904$\pm$92  & 9.69$\pm$2.43\\
80    &    248$\pm$19 &  914$\pm$60  &11.83$\pm$4.71\\
90    &    191$\pm$12 &  994$\pm$82  & 6.19$\pm$1.54\\
100   &    180$\pm$22 &  754$\pm$65  &12.97$\pm$3.18\\
110   &    269$\pm$25 &  922$\pm$139 &11.03$\pm$2.76\\
120   &    211$\pm$20 &  933$\pm$68  &11.61$\pm$3.40\\
130   &    140$\pm$18 &  807$\pm$57  & 7.79$\pm$1.94\\
140   &    165$\pm$13 &  949$\pm$65  & 8.34$\pm$2.08\\
150   &    221$\pm$16 &  827$\pm$67  &13.97$\pm$3.51\\
160   &    167$\pm$17 & 848$\pm$71  & 11.26$\pm$3.90\\
170   &    214$\pm$14 &  872$\pm$58  &12.52$\pm$3.89\\
180   &    183$\pm$14 &  839$\pm$54  &12.24$\pm$3.07\\
190   &    143$\pm$18 &  815$\pm$33  & 9.99$\pm$3.48\\
200   &    172$\pm$19 &  807$\pm$51  &12.81$\pm$3.94\\
210   &    186$\pm$13 &  956$\pm$66  &10.31$\pm$2.59\\
220   &    160$\pm$17 &  959$\pm$103 & 9.32$\pm$2.34\\
230   &    199$\pm$16 & 1008$\pm$68  &11.59$\pm$2.59\\
240   &    201$\pm$21 &  922$\pm$58  &13.84$\pm$3.46\\
250   &    176$\pm$20 &  903$\pm$80  &11.93$\pm$4.72\\
260   &    210$\pm$11 &  965$\pm$89  & 7.21$\pm$1.81\\
270   &    225$\pm$22 & 1003$\pm$36  & 8.83$\pm$1.52\\
280   &    197$\pm$20 & 1002$\pm$65  & 8.07$\pm$2.05\\
290   &    192$\pm$28 & 1005$\pm$120  & 9.75$\pm$3.46\\
300   &    198$\pm$23 &  991$\pm$129 &13.59$\pm$3.40\\
310   &    209$\pm$15 & 1149$\pm$89  &12.62$\pm$3.16\\
320   &    176$\pm$17 & 1155$\pm$106 &11.09$\pm$2.85\\
330   &    214$\pm$16 & 1096$\pm$71  &13.76$\pm$4.44\\
340   &    149$\pm$24 &  987$\pm$57  & 9.70$\pm$3.47\\
350   &    174$\pm$21 & 1066$\pm$80  &10.94$\pm$2.74\\
\hline
\end{tabular} 
\end{table}

\section{Dependence of the Galactic model parameters with the Galactic longitude}
\begin{figure}[!h]
\begin{center}
\includegraphics[angle=0, width=70mm, height=55.2mm]{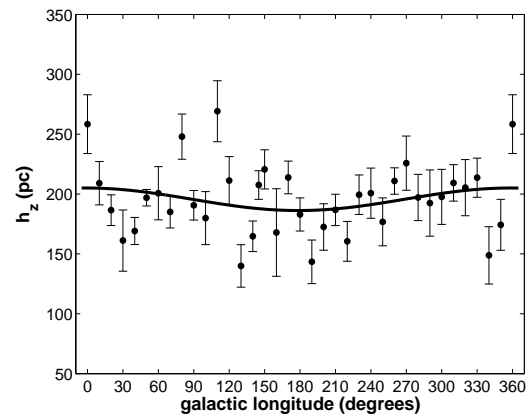}
\caption[] {Variation of the scaleheight of the thin disc with the Galactic longitude.} 
\label{hz-TN-L}
\end{center}
\end {figure}

Fig. \ref{hz-TN-L} shows the variation of 
the scaleheight of the thin disc with the Galactic longitude. No obvious global trend is observed in the thin disc scaleheights, 
with a sample of values well compatible with a constant value of $<h_z>$=187 ($\sigma$=36) pc, well in agreement with 
usual estimates for this component \citep[e.g.,][]{Bahcall84, Robin86, RM93, Siegel02, L02}. As shown in Fig. \ref{hz-TN-L}, we 
have overplotted a sinusoidal fit, no physical meaning, to the data. One can say that the scaleheight of the thin disc shows slight 
variation with the Galactic longitude. However, the scatter and the error bars are large, being compatible with a constant value 
independent of the Galactic longitude. Hence, we can conclude that the thin disc scaleheight is found to be a constant as a function
of Galactic longitude.

\begin{figure}[!h]
\begin{center}
\includegraphics[angle=0, width=70mm, height=55.2mm]{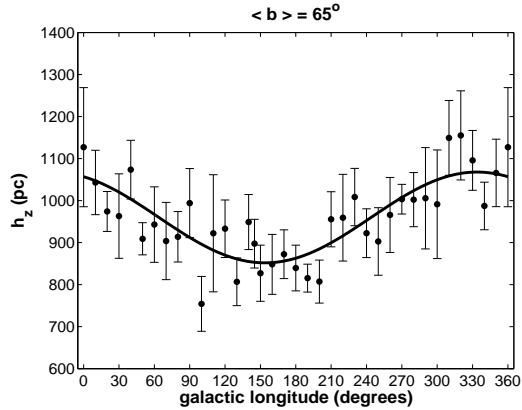}
\caption[] {Variation of the scaleheight of the thick disc with the Galactic longitude.}
\label{hz-TK-L}
\end{center}
\end {figure}

For the thick disc the variation of the scaleheight with the Galactic longitude is more pronounced (Fig. \ref{hz-TK-L}). There is a
maximum in the inner Galaxy ($|l|<30^{\circ}$), whereas in 
between $150^{\circ}<l<230^{\circ}$ there is a local minimum. The mean 
of the scaleheight for the data, \mbox{$<h_z>=957$} ($\sigma$=100) pc  is well 
within the given range found in the literature \citep{Spagna96, Ng97, Buser98, Buser99, Siegel02, Larsen03, Cabrera05}.

\begin{figure}
\begin{center}
\includegraphics[angle=0, width=70mm, height=55.2mm]{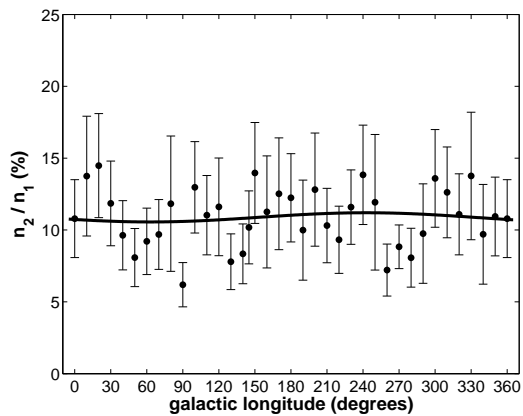}
\caption[] {Variation of the relative space density of the thick disc with the Galactic longitude.} 
\label{thick_disc_den}
\end{center}
\end {figure}

The variation of the local space density of the thick disc relative to the local space density of 
the thin disc ($n_{2}/n_{1}$) with the Galactic longitude is given in Fig. \ref{thick_disc_den}. The trend gives a constant local 
space density, $(n_{2}/n_{1})\sim 10$ per cent, which confirms that the variation in $h_z$ in Fig. \ref{hz-TK-L} is not an artifact of an
anticorrelation in the local normalization (assuming, for example, a constant value $(n_{2}/n_{1})=10\%$ we still get the result in 
Fig. \ref{hz-TK-L}).

\begin{figure}
\begin{center}
\includegraphics[angle=0, width=70mm, height=55.2mm]{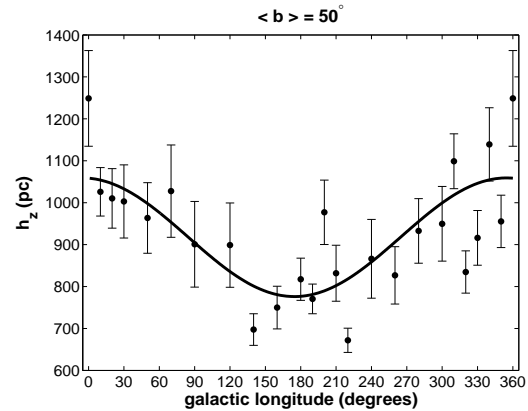}
\caption[] {Variation of the scaleheight of the thick disc with the Galactic longitude, for fields with Galactic latitude $<b>=50^\circ$.}
\label{hz_thick_50}
\end{center}
\end {figure}
       
We checked the effect of the Galactic latitude on the variation of the scaleheight of the thick disc with the Galactic longitude for the {\em
2MASS} data. Fig. \ref{hz_thick_50} shows the variation of the scaleheight with the Galactic longitude for 
fields at $<b>=50^{\circ}$. The trends in Figs. \ref{hz-TK-L} and \ref{hz_thick_50} are similar, 
although the zero points are different. The equations of the sinusoidal fits for the data in the fields with 
$<b>=50^{\circ}$ and $<b>=65^{\circ}$ are as follows:  
\begin{equation}
h_z = 918 - 141.45~ \sin ~ (l-84.65^\circ), ~~~~~~(at <b>=50^\circ),
\label{senos1}
\end{equation}
\begin{equation}
%h_z = 956 - 97.57~ sin (l-68.21^\circ), ~~~~~~~(at <b>=65^\circ).
h_z = 959 - 108.02~ \sin ~ (l-63.87^\circ), ~~~~~~~(at <b>=65^\circ).
\label{senos2}
\end{equation}
The difference between the scaleheights is larger (it amounts up to $\sim 100$ pc) for the longitudes corresponding to different minima in the
two figures. No evident differences are found between the overall shape of the distributions. Hence the observed behaviour is associated to 
a global trend in the thick disc rather than to something exclusive of a restricted range of latitudes.

\section{Discussion}

\subsection{Does the flare affect the scaleheight?}

It is well known  that the Galactic disc shows a flare, which  produces an increase in the scaleheight as we move outwards in the Galaxy. In \cite{L02} the flare was modelled by an exponential increase of $h_z$ with $R$, obtaining a law that approximately reproduced the observed counts up to $R<15$ kpc. On the contrary, for the inner Galaxy a flare with the opposite trend (an increase in $h_z$ as $R$ decreases) was found by \cite{L04}, who then proposed an expression that summarizes both regimes with a smooth transition between them:

\begin{equation} 
h_z (R) = h_z (R_\odot) \times [1 + 0.21 (R-R_\odot)+ 0.056 (R-R_\odot)^2]
\end{equation}

\begin{figure}
\begin{center}
\includegraphics[angle=0, width=70mm, height=55.2mm]{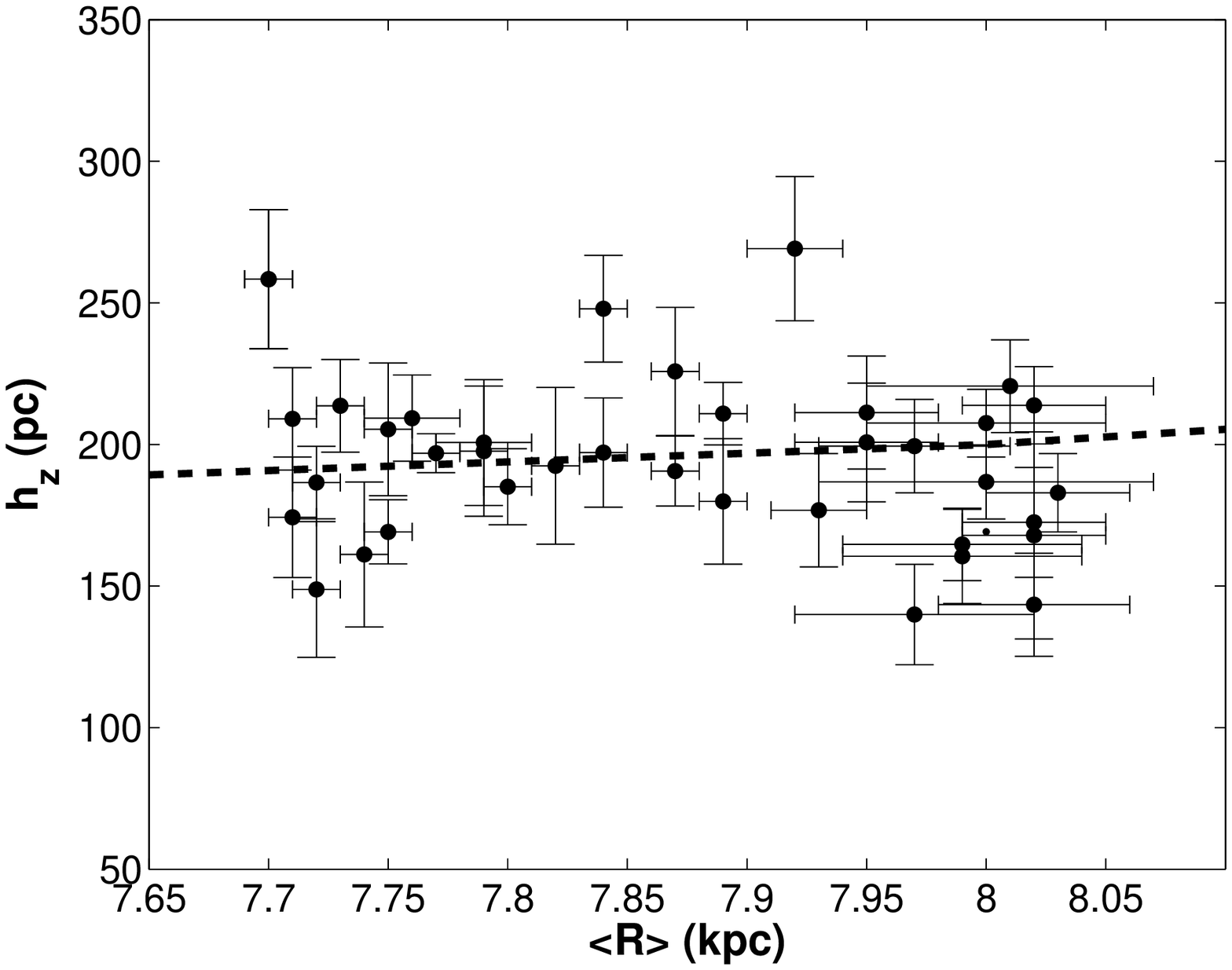}
\includegraphics[angle=0, width=70mm, height=55.2mm]{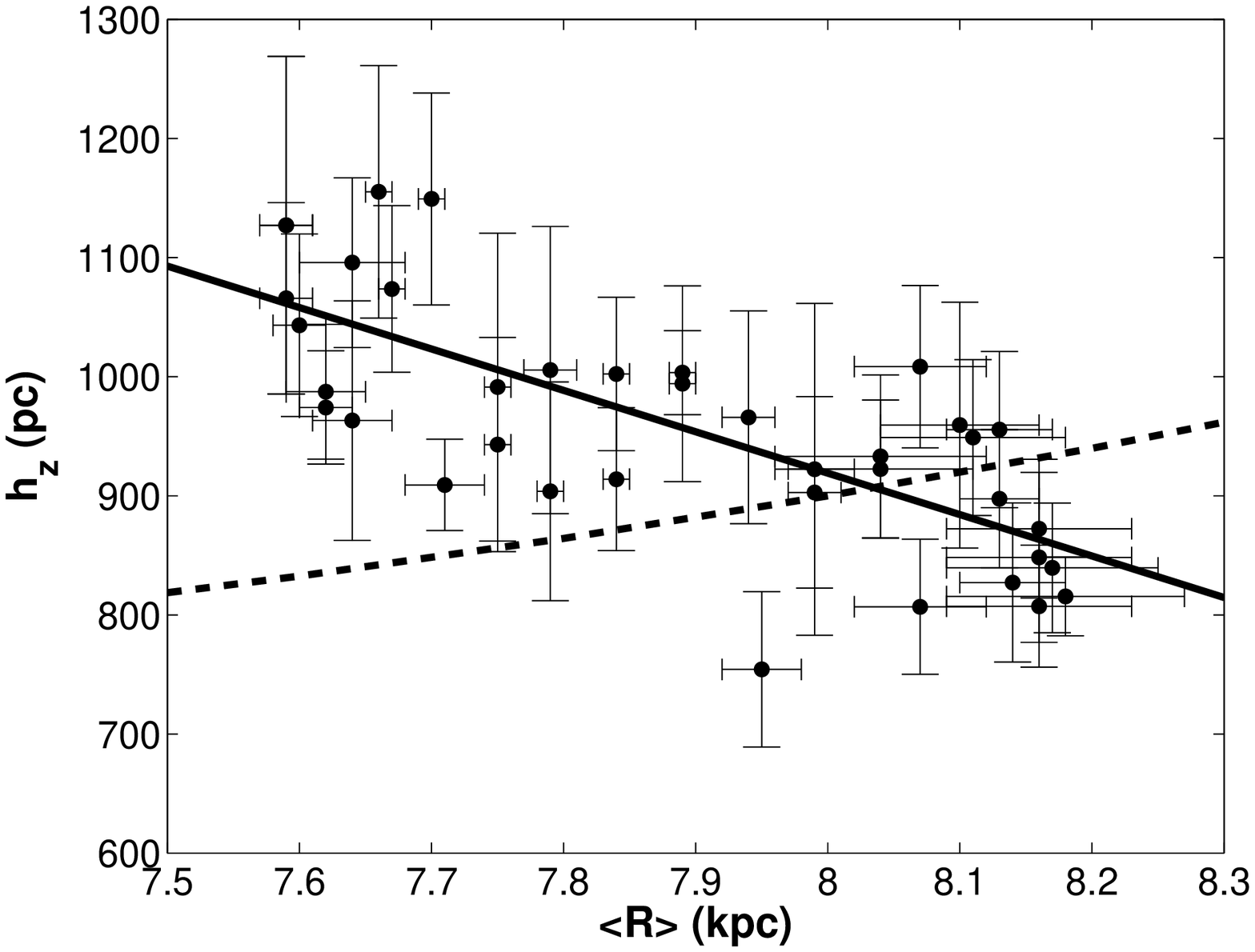}
\caption[] {Variation in the scaleheight of the thin (above) and thick (below) discs with $R$.} 
\label{flare}
\end{center}
\end {figure}

The variation of the scaleheight of the thin and thick disc with Galactic longitude could be related with the flare itself. 
When the line of sight is pointing to the inner Galaxy ($|l|<30^\circ$) the mean Galactocentric distance of the sources is 
lower than when the line of sight is in the anticentre direction. We can then translate the $h_z$ vs.\ longitude plot in a 
$h_z$ vs.\ $R$ plot by deriving the mean Galactocentric distance to either the thin or thick disc sources, a procedure that is made
by binning the data by the Galactocentric distance, R, and determining the mean radii of the red clump stars in each 
bin to average them. Figure \ref{flare} shows the variation in the scaleheight of the thin and thick disc with Galactocentric distance. It has to be 
noted that the range of Galactocentric distances is very small, so a possible derivation of parameters for the flare is far
 from being conclusive. We have compared L\'opez-Corredoira et al.'s law in both cases 
 (dashed lines in the plot) considering $h_z$ ($R_\odot)=200$ pc and $h_z$ $(R_\odot)=900$ pc for the thin and thick 
 disc, respectively. While the predictions for the thin disc are compatible with the data (although a 
 constant value of $h_z$ can also fit the data), for the thick disc the observed trend is just the opposite 
 with an increase of $h_z$ when moving to lower values of $R$ (represented by means of a linear fit 
 on the lower panel of Fig. \ref{flare}), a result that is in complete disagreement with the obtained for the outer 
 thin disc (R$>$6 kpc) in L\'opez-Corredoira et al. (2002).
 
\begin{figure}
\begin{center}
\includegraphics[angle=0, width=70mm, height=55.2mm]{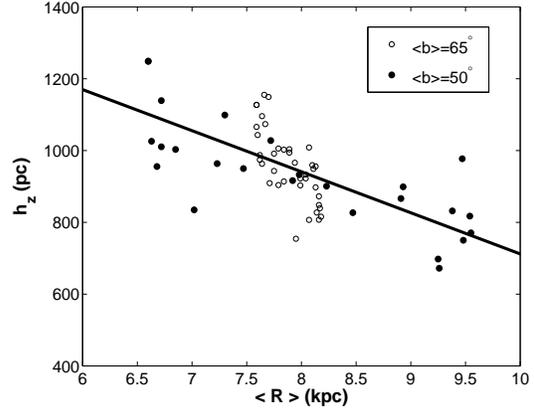}
\caption[] {Variation of the scaleheight of the thick disc with $R$ by using {\em 2MASS} data at 
$<b>=50^\circ$ and $<b>=65^\circ$. The best linear
fit to the observed trend is also shown.} 
\label{thickflare}
\end{center}
\end {figure}

To obtain a more accurate representation of the flare in the thick disc it would be necessary to increase the range
 of Galactocentric distances. For this reason, {\em 2MASS} data at $<b>=50^\circ$ are useful. They 
 correspond to lines of sight at lower heights above the plane, thus the mean
Galactocentric distance of the thick disc sources increases improving the fit. In Fig. \ref{thickflare}
 we show the variation in the scaleheight of the thick disc with respect to the mean Galactocentric distance, combining the 
 results at $<b>=50^\circ$ (filled circles) with those at $<b>=65^\circ$ (open circles). By means of 
 a linear fit to the data we obtain the following expression:

\begin{equation} 
h_z (R)= (940\pm20)\times [1 - (0.12\pm0.02) (R-R_\odot)] ~~~~~~ (pc)
\label{hzthick}
\end{equation}

It is not so common to observe an increase of the scaleheight as we move to the inner Galaxy. Furthermore, 
no similar analysis of a possible flare in the thick disc is found in the literature. In a very recent article, 
\cite{Momany06}  analysed the warp and flare of the disc of the Milky Way but only concentrated in those related to the thin disc. 
By using red giant branch stars in the range $0.45 < z < 2.25$ kpc they obtained a slight increase of the scaleheight of the disc with increasing Galactocentric radii, up to $R\sim$15 kpc (their fig.\ 15). They claimed that the overall mean scaleheight around 0.65 kpc they obtained was a reflection of a mixture of the thin and thick disc populations. If we put together the known trend in the scaleheight of the thin disc (that is, an increasing $h_z$ as $R$ increases) with the result obtained in this paper for the thick disc (an increase in $h_z$ as $R$ decreases), one must expect that a sort of 'cancellation' appears, producing a nearly constant $h_z$ at intermediate 
heights above the plane.

To check this possibility, we have made a simple simulation for a combination of a thin disc 
with the density laws of \cite{L04} and a thick disc with an increase in $h_z$ represented by eq. (\ref{hzthick}) and a 
relative space density of 10 per cent of the thin disc according to \cite{Cabrera05}. The resulting densities are fitted by a single exponential law (as if only one component were present), obtaining in that way the expected variation in scaleheight with the Galactocentric distance. 

\begin{figure}
\begin{center}
\includegraphics[angle=0, width=80mm, height=65.2mm]{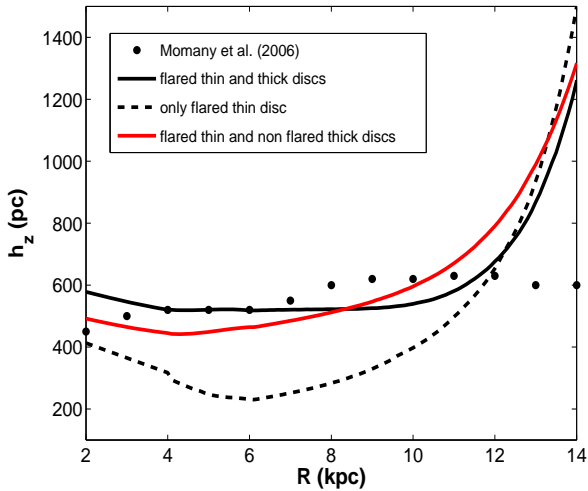}
\caption[] {Variation of the scaleheight with $R$ taken from the work of \cite{Momany06} compared with the 
simulations of different possible
combinations of thin and thick disc densities, following the results of \cite{L04} (for the thin disc) and 
this paper (for the thick disc).} 
\label{momany}
\end{center}
\end {figure}

When comparing this simulation with the data taken from \cite{Momany06} (Fig. \ref{momany}) a good agreement is observed. In the range of
distances  $4 < R < 9$ kpc  a nearly constant scaleheight is expected that is not far from that observed (note we did not perform any kind
of fit to Momany et al.'s data, so the coincidence in the value for $h_z$ at $4 < R < 7$ kpc is very in\-te\-res\-ting). Predicted values for $R>10$ kpc are
larger than those, but it was expected due to the excessively high increase in $h_z$ for the thin disc that L\'opez-Corredoira et al.'s model produces in this range of Galactocentric distances. 
A more modest increase in $h_z$ with $R$ for the thin disc (as obtained in \cite{Al00} or in \cite{Momany06}) would produce a better agreement with the data, but this was beyond the scope of this comparison. For $R<4$ kpc, again an excessive increase in $h_z$ is obtained in the simulation, but again it was known that the validity of L\'opez-Corredoira et al.'s model for $R<2.5$ kpc is not well demonstrated. 

We have also introduced some changes in the simulation. By eliminating the contribution of the thick disc (that is, assuming that only the thin disc is observed) we recovered the predictions of the L\'opez-Corredoira et al.'s model, with the scaleheight corresponding to the thin disc (dashed line in Fig. \ref{momany}). It is even more remarkable how when a flaring in the thick disc is neglected (assuming a constant scaleheight of 940 pc for this component) the trend obtained still follows the increase in the scaleheight due the thin disc, but with higher $h_z$ values as corresponds with the mixing of both populations (grey solid line in Fig. \ref{momany}). The result is also less sensitive to other parameters, as the relative space density of the thick disc (no changes were obtained by varying it from 5 to 15 per cent) or the value of $R_\odot$ (assumed as 8 kpc).

A combination of a flared thin disc and a flared thick disc with opposite trends seems, then, to be compatible with the observed scaleheights at
intermediate heights above the plane, as those of \cite{Momany06}. Thus, the observed variation in the scaleheight of the thick disc with the galactic longitude has been obtained here as expected.

\subsection{Effect of the scalelength of the thick disc}
\label{seccH}
In Section \ref{2mass} we addressed what we considered a fixed
scalelength of 3 kpc for the thick disc, as obtained in
\cite{Cabrera05}. However, a different scalelength would reproduce the
observed variation in the scaleheight without the need for a flare in the
 disc itself. For example, a shorter scalelength of 2 kpc for the thick disc
would increase the
predicted star counts in the inner disc and reduce them in the outer disc,
relative to the assumed 3 kpc scalelength, simulating a
higher scaleheight in the inner disc and a lower scaleheight in the outer
disc, as  was found in Section 4.

However, the effect of the 
scalelength on the derived densities is negligible, as we are moving in a
very narrow range of Galactocentric distances. To
reproduce the same density as that resulting from a scalelength of 3 kpc
and a scaleheight of 1100 pc, like those derived in the inner Galaxy but
with a constant scaleheight of 800 pc, the scalelength has to change by a
factor $\sim$1.4 with respect to the assumed value of 3 kpc (that is,
around 4.2 kpc), a value that is 11$\sigma$ above that derived in
\cite{Cabrera05}.

\begin{figure}[!h]
\begin{center}
\includegraphics[angle=0, width=70mm, height=55.2mm]{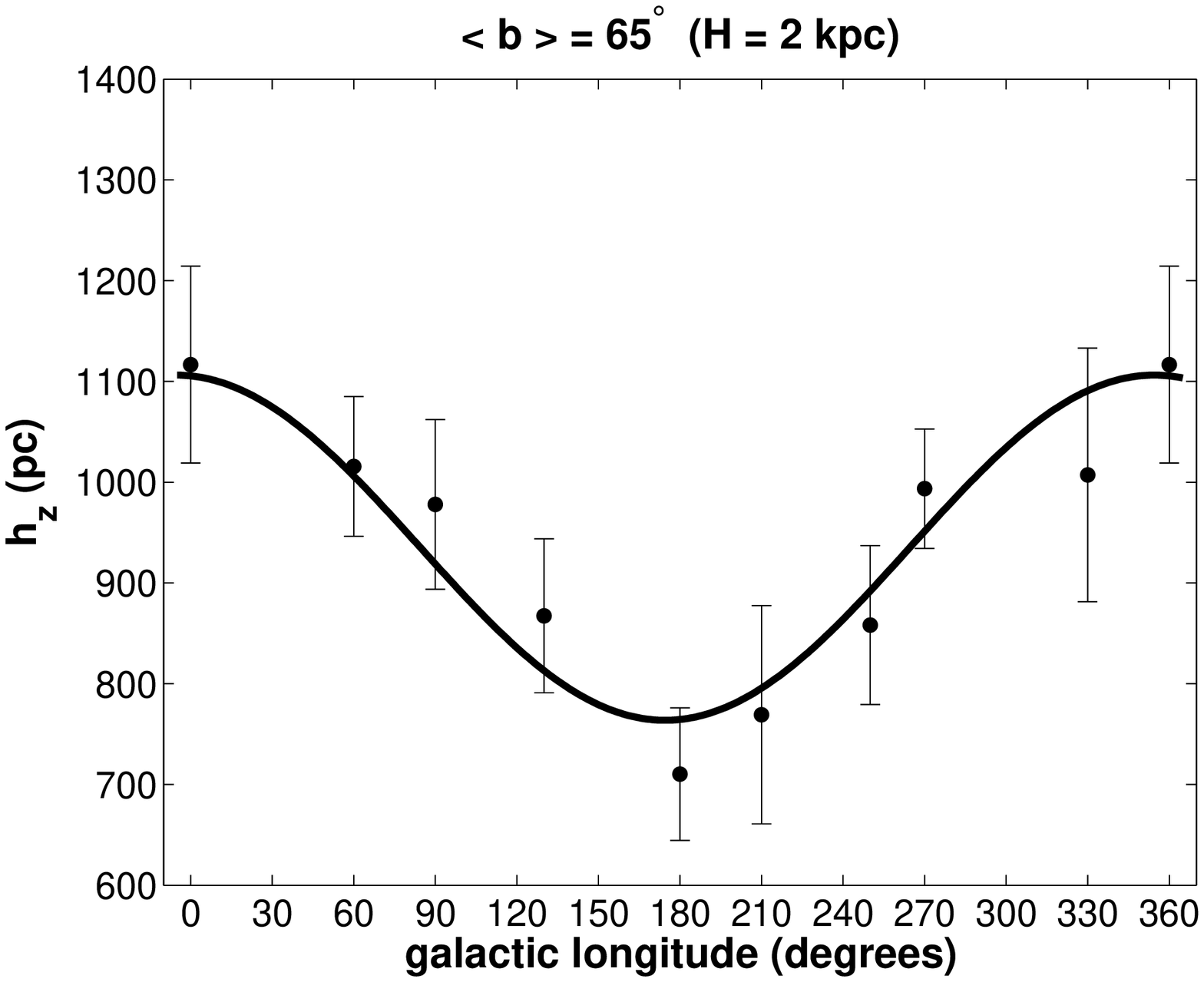}
\includegraphics[angle=0, width=70mm, height=55.2mm]{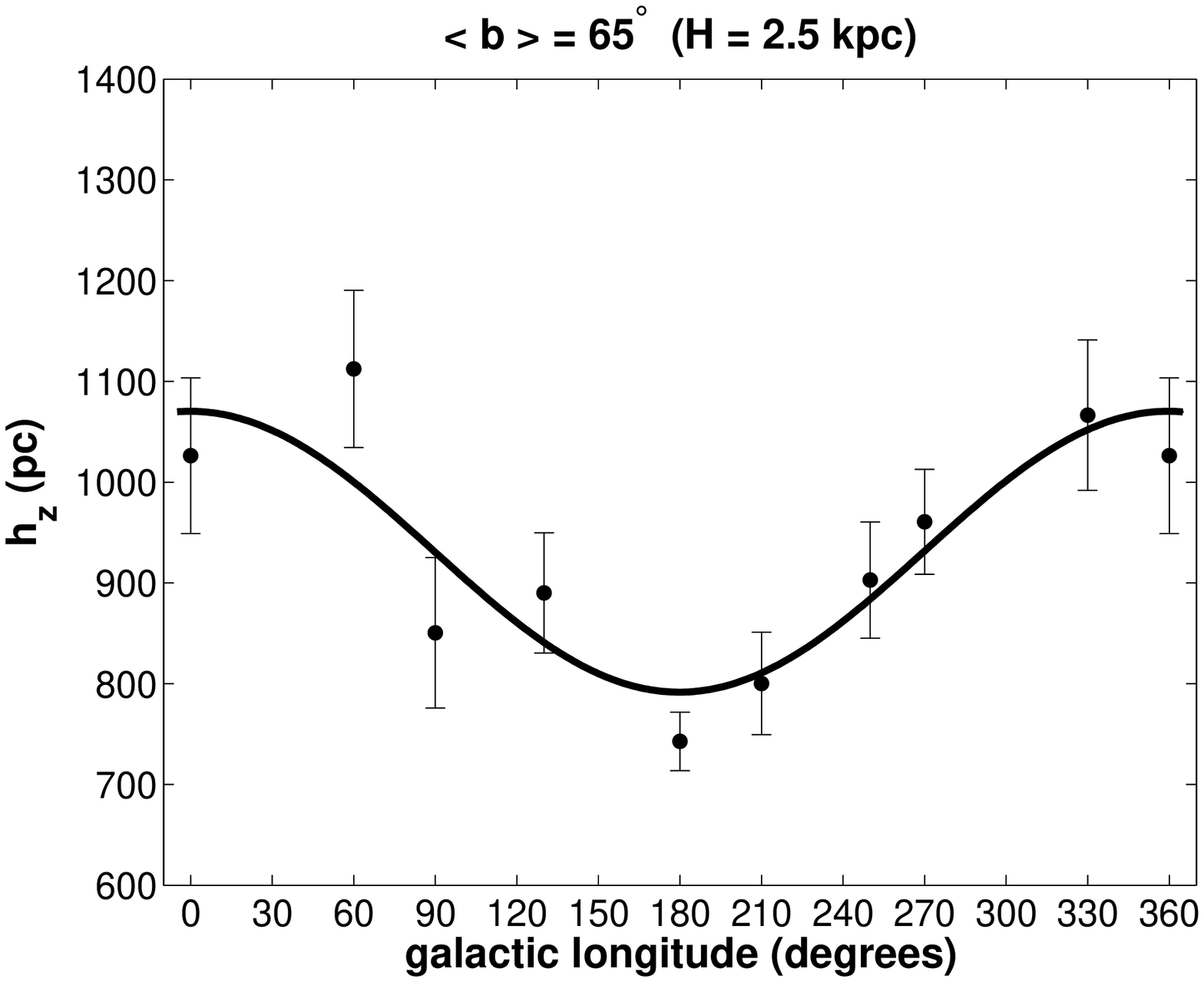}
\caption[] {Variation in the scaleheight of the thick disc with the Galactic longitude obtained by assuming a scalelength of $H$=2 kpc (above) 
or $H$=2.5 kpc for this component (below).} 
\label{Hthick}
\end{center}
\end {figure}

In order to check this statement, we repeated the
analysis of Section 2 in a series of nine test fields but now assuming two 
different scalelengths for the thick disc: $H$ = 2 kpc and $H$ = 2.5 kpc,
respectively. The results are shown in Fig. \ref{Hthick}, together with sinusoidal
fits to the data. The observed trend is nearly coincident with that of
Fig. \ref{hz-TK-L} in both  amplitude and the positions of the minima. 
 Hence, the effect of a different scalelength  from the assumed one of 3
kpc has to be discarded as being responsible for the results discussed in
 this paper.

\section{Conclusion}
By using NIR data from  {\em 2MASS} we have observed significant changes
in the 
structural parameters of the thin and thick discs of the Milky Way, that
 are more than simple fluctuations caused by model fitting to the data. 

A plausible explanation for the observed changes in the scaleheight of the
discs arises from the combined effect of the warp and flare in the
Galactic thin and thick disc, which displaces both structures from the
assumed position expected for a ``smooth disc'' (that is, a flat, azimuthally homogeneus and constant
scaleheight disc). This implies that in some 
directions the effect is more noticeable than in others, producing
variations in the parameters resulting from a fit of a single Galactic model. 
 The observed trend in the change of the scaleheight with galactic
longitude is well reproducible theoretically by assuming the accretion of 
intergalactic matter on to the Galactic disc, which in general
produces a different pressure depending on both the Galactocentric radius 
and azimuth, the same pressure which would also be responsible for the
formation 
of S-warps and/or U-warps in  galaxies \citep{LBB02}. The pressure
would be similar to a piston mechanism, 
only from one side of the disc \citep[][\S 4.6]{Salcedo06}. As a result of
this mechanism the scaleheight of
 the disc would present a similar shape to that  obtained in Figure
\ref{hz-TK-L}. However, it is far from our 
 intention to consider that our result is a demonstration that this 
speculative hypothesis is the correct one in explaining the uncertain
origin for the Milky Way's flare and warp; but it is consistent with the observed data. 

The flare obtained in the thick disc from {\em 2MASS} data has an opposite
trend to that commonly assumed for the thin disc, 
with a scaleheight that increases as we move to the innermost Galaxy,
being the first time that a 
possible flare in this Galactic component is observed. It is beyond the scope of this paper to determine 
from the observed results a
theoretical scenario that supports the origin of this possible flare,
although some asymmetries in 
the thick disc hve been well known for some time now \citep{Parker03}.
We suspect that the different behaviour in the inner Galaxy is related to
the in\-te\-rac\-tion of the
4 kpc inner Galactic bar \citep{L06} with its surroundings.  However, it is
more difficult to explain
the possible effect of this component on the thick disc, although some known
asymmetries of this component have been suggested as having been caused by the
Galactic bar \citep{Parker04}. Unfortunately, we do not have  enough
information to add anything in this regard. Probably, kinematic
analyses by using radial velocity data as those provided by the \emph{Sloan Extension for Galactic 
Understanding and Exploration} (SEGUE, Newberg et al., 2003) or the \emph{Radial Velocity Experiment} (RAVE, Steinmetz et al., 2006) 
will be very valuable to go deeper in this topic.

\begin{acknowledgements}
Thanks are given to Dr. Chris Flynn, who made very important suggestions
that have improved the overall quality of the work presented in this paper.
We also thank Dr. Salih Karaali for many comments and discussions during the
preparation of this paper. This work was supported by the Research Fund of
the University of Istanbul, with project number BYPF-11-1.

This publication makes use of data products from the Two Micron All Sky Survey, which is a joint project of the University of Massachusetts and the Infrared Processing and Analysis Center/California Institute of Technology, funded by the National Aeronautics and Space Administration and the National Science Foundation. 
\end{acknowledgements}

\end{document}